# METAL HYDROGEN SULFIDE CRITICAL TEMPERATURE AT THE PRESSURE 225 GPa

**N. A. Kudryashov, A.A. Кutukov, E. A. Mazur** [a)]

National Research Nuclear University "MEPHI", Kashirskoe sh.31, Moscow 115409, Russia

Éliashberg theory, being generalized to the electron-phonon (EP) systems with not constant density of electronic states, as well as to the frequency behavior of the renormalization of both the electron mass and the chemical potential, is used to study hydrogen sulphide $SH_3$ phase $T_c$ under pressure. The phonon contribution to the anomalous electron Green's function (GF) is considered. The pairing is considered within the overall width of the electron band, and not only in a narrow layer at the Fermi surface. The frequency and the temperature dependences of the electron mass renormalization $\mathrm{Re}Z(\omega)$, the density of electronic states $N(\varepsilon)$, renormalized by the EP interaction , the spectral function of the electron-phonon interaction, obtained by calculation are used to calculate the electronic abnormal GF. The Éliashberg generalized equations with the variable density of electronic states are resolved. The dependence of both the real and the imaginary part of the order parameter on the frequency in the $SH_3$ phase is obtained. The $T_c = 177K$ value in the hydrogen sulfide $SH_3$ phase at the pressure P = 225 GPa has been defined.

## 1. The theory of the superconductivity for the e-band with the not constant density of electronic states

The calculation of the superconducting transition temperature in the hydrogen sulphide [1,2] is usually carried out either by using extremely lax and insufficiently reasoned description of superconductivity with a freely selectable electron density functional [3–5] without any possibility of taking into account the effect of nonlinear nature of Éliashberg equations either within the approximate solutions of the Éliashberg equations [6–8] not taking into account the variable nature of the electronic density of states $N(\omega)$. The solution of the Éliashberg equations on the real axis is considered to be a difficult task. Therefore the critical temperature $T_c$ with the Éliashberg formalism is usually calculated in the Green's function presentation with the use of a set of discrete Green's function values on the imaginary axis and not taking into account the variable nature of the electronic density of states (see [9] for such calculation in the hydrogen sulfide).The analytic continuation of such a solution in order to determine the frequency dependence of the order parameter is extremely inaccurate. The aim of present work is to derive the generalized Éliashberg equations for the precise $T_c$ calculation in the materials with the strong EP interaction, allowing a quantitative calculation and prediction of the superconducting properties and $T_c$ in the different phases of the hydrogen sulfide [1,2], as well as in the high-temperature materials with the EP mechanism of superconductivity, which will be discovered in the near future. We will take into account all the features of the frequency behavior of the spectral function of the EP interaction , the behavior of the electron density of states $N_0(\omega)$ and the specific properties of matter in which the superconducting state is established.

For this purpose we construct in the present paper a revised version of the Migdal-Éliashberg theory for the EP system at nonzero $T \neq 0$ temperature in the Nambu representation with the account of several factors such as the variability of the electronic density of states $N_0(\varepsilon)$ within the band, the frequency and temperature dependence of the complex mass renormalization $\operatorname{Re} Z(\omega,T)$, $\operatorname{Im} Z(\omega,T)$, the frequency and temperature dependence of the $\operatorname{Re}\chi(\omega,T)$, $\operatorname{Im}\chi(\omega,T)$ terms usually referred to as the «complex renormalization of the chemical potential», the spectral function of the electron-phonon interaction obtained by calculation, as well as the effects resulting from both the electron-hole non-equivalence and the fact of the zone width finiteness.

Under these conditions, the derivation of the Éliashberg equations [10–11] is performed anew on a more rigorous basis, leading to the new terms in the equations for the order parameter, which are not accounted for in the previous versions of the theory (see., e.g. [12–24]). As it has been shown in [25] in the case of the strong electron-phonon interaction the reconstruction of the real part $\mathrm{Re}\Sigma$ as well as of the imaginary part $\mathrm{Im}\Sigma$ of the GF self-energy part (SP) in the materials with the variable density of electronic states is not limited to the frequency $\omega$ domain $\omega_D$ of the limiting phonon frequency, and extends into the much larger frequency range $\omega \gg \omega_D$ from the Fermi surface. As a result, the EP interaction modifies SP of the Green's function, including its anomalous part, at a considerable distance from the Fermi surface in the units of Debye phonon frequency $\omega_D$, and not only in the vicinity $\mu - \omega_D < \omega < \mu + \omega_D$ of the Fermi surface.

Given all that is written above, we consider EP system with the Hamiltonian which includes the electronic component $\hat{H}_e$, the ionic component $\hat{H}_i$ and the component corresponding to the electron-ion interaction in the harmonic approximation $\hat{H}_{e-i}$, so that $\hat{H} = \hat{H}_e + \hat{H}_i + \hat{H}_{e-i} - \mu\hat{N}$. Here the following notations are introduced: $\mu$ as the chemical potential, $\hat{N}$ as the number of electrons operator in the system. The matrix electron Green function $\hat{G}$ is defined in the representation given by Nambu as follows $\hat{G}(x,x') = -\langle T\Psi(x)\Psi^+(x')\rangle$, where conventional creation and annihilation operators of electrons appear as Nambu operators. SP of the retarded GF at the discrete frequency points set $\omega_m = (2m+1)\pi T, m = 0, \pm 1, \pm 2..$ on the imaginary axis can be written as $\hat{\Sigma}(i\omega_m) = i\omega_m\left[1 - Z(\vec{p},\omega_m)\right]\hat{\tau}_0 + \chi(\vec{p},\omega_m)\hat{\tau}_3$. $\chi(\xi,\omega)$ is the complex function, the real part of which is commonly referred to as the EP renormalization of the chemical potential interaction. The real part of the $\chi(\vec{p},\omega_m)$ term after analytic continuation determines the frequency-dependent shift of the chemical potential given by the following formula $\chi(\vec{p},\omega_m) = \frac{1}{2}\left[\Sigma(\vec{p},\omega_m) + \Sigma(\vec{p},-\omega_m)\right]$. The $Z(\vec{p},\omega_m)$ term the real part of which after analytic continuation sets the renormalization of the electron mass, and the imaginary part sets the damping of an electron is given by the expression $i\omega_m\left[1-Z\left(\vec{p},\omega_m\right)\right]=\frac{1}{2}\left[\Sigma\left(\vec{p},\omega_m\right)-\Sigma\left(\vec{p},-\omega_m\right)\right]$. After analytical continuing of the terms $Z\left(\vec{p},i\omega_m\right)$ and $\chi\left(\vec{p},\omega_m\right)$ the emerging functions $\operatorname{Re}Z\left(\vec{p},\omega\right)$ and $\operatorname{Re}\chi\left(\vec{p},\omega\right)$ become even for all frequency values $\omega$, including the frequency values at the real axis. After analytic continuation $i\omega_p\rightarrow\omega+i\delta$ the phonon contribution to the self energy part of the retarded e-GF $\hat{g}_R$ is expressed as follows:

$$\hat{\Sigma}^{ph}(\xi,\omega)=\frac{1}{\pi}\int_{-\infty}^{+\infty}dz'\int_{-\mu}^{+\infty}d\xi'\frac{N_0\left(\xi'\right)}{N_0\left(0\right)}K^{ph}(z',\omega)\hat{\tau}_3\operatorname{Im}\hat{g}_R(\xi',z')\hat{\tau}_3. \tag{1}$$

We use the technique for the real frequency Éliashberg equations' solution. This technique will allow us to control the process of calculating $\operatorname{Re}Z\left(\omega\right)$, $\operatorname{Im}Z\left(\omega\right)$, $\operatorname{Re}\Sigma\left(\omega\right)$, $\operatorname{Im}\Sigma\left(\omega\right)$, $\operatorname{Re}\chi\left(\omega\right)$ and $\operatorname{Im}\chi\left(\omega\right)$ frequency behavior. In (1) the Pauli matrices $\hat{\tau}_i$ are introduced, $\alpha^2F$ is the EP interaction spectral function, $N_0\left(\xi\right)$ is a "bare" (not renormalized with the EP interaction) variable electronic density of states defined by the following expression $\int\limits_{S(\xi)}\frac{d^2\boldsymbol{p}'}{v_{\xi\boldsymbol{p}'}}d\xi=\int\limits_{S(\xi)}N_0\left(\xi\right)d\xi$ with the energy of the "bare" electrons $\xi$ with the pulse $\boldsymbol{p}$ measured from the Fermi level. It is not assumed that the electron pulses are on the Fermi surface. Neglect in (1) $\alpha^2F$ dependence on the $\xi$, $\xi'$ variables so that $\alpha^2\left(\xi',\xi,z\right)F\left(\xi',\xi,z\right)\approx\alpha^2\left(z\right)F\left(z\right)$. Replace the $Z\left(\vec{p}',\omega\right)$ function with $Z\left(\omega\right)$ corresponding to the constant energy $\xi$ in the direction determined by the angle $\varphi$. Average the expression (1) to the angle $\varphi$ of the pulse direction. In the transition from the integration $\int_{-\infty}^{+\infty}dz'$ to the integration $\int_{0}^{+\infty}dz'$ take into account the $\operatorname{Re}Z\left(z'\right)$ parity, as well as the following order parameter property $\varphi(-z')=\varphi^*\left(-z'\right)$ [ 17]. From (1), taking into account the standard expression for the retarded Green's function $\hat{g}_R(\xi',z')$ we obtain the equations for both the real $\mathrm{Re}\,\varphi(\omega)$ and imaginary part $\mathrm{Im}\,\varphi(\omega)$ of the anomalous GF self-energy part as the system of two equations having the following form:

$$\mathrm{Re}\,\varphi(\omega) = -\frac{1}{\pi} P \int_0^{+\infty} dz' \left[ K^{ph}(z',\omega) - K^{ph}(-z',\omega) \right] \int_{-\mu}^{+\infty} d\xi' \frac{N_0(\xi')}{N_0(0)} \mathrm{Im} \frac{\varphi(z')}{\left[ Z(z')(z') \right]^2 - \varphi^2(z') - \left( \xi' + \chi(z') \right)^2},$$

(2)

$$\mathrm{Im}\,\varphi(\omega) = -\frac{1}{2} \int_0^{+\infty} dz' \{ \alpha^2(|\omega - z'|) F(|\omega - z'|) \left[ cth\frac{(\omega - z')}{2\mathrm{T}} + th\frac{z'}{2\mathrm{T}} \right] sign(\omega - z') -$$
$$-\alpha^2(|\omega + z'|) F(|\omega + z'|) \left[ cth\frac{(\omega + z')}{2\mathrm{T}} - th\frac{z'}{2\mathrm{T}} \right] sign(\omega + z') \} \times \qquad (3)$$
$$\times \int_{-\mu}^{+\infty} d\xi' \frac{N_0(\xi')}{N_0(0)} \mathrm{Im} \frac{\varphi(z')}{\left[ Z(z')(z') \right]^2 - \varphi^2(z') - \left( \xi' + \chi(z') \right)^2}$$

where

$$K^{ph}(z',\omega) = \int_0^{+\infty} dz \alpha^2(z) F(z) \frac{1}{2} \left\{ \frac{th\frac{z'}{2\mathrm{T}} + cth\frac{z}{2\mathrm{T}}}{z' + z - \omega} - \frac{th\frac{z'}{2\mathrm{T}} - cth\frac{z}{2\mathrm{T}}}{z' - z - \omega} \right\}. \qquad (4)$$

In (2-3) we have neglected in the first approximation the Coulomb contribution to the order parameter in view of the smallness of the Coulomb pseudopotential $\mu^* \approx 0.1$ in the hydrogen sulfide SH3 phase compared with the significant constant $\lambda \sim 2.21$ of the EP interaction in the sulfide phase. The expressions and the graphics for the $(1,1)$ component of the imaginary part of the SP matrix $\operatorname{Im}\Sigma(\omega) = -\operatorname{Im}Z(\omega)\omega + \operatorname{Im}\chi(\omega)$ near the $T_c$ midrange were obtained in [27]. By direct calculation we get the following expression appearing in (2), (3):

$$\operatorname{Im}\frac{\varphi(z')}{\left[Z(z')(z')\right]^2 - \varphi^2(z') - \left(\xi' + \chi(z')\right)^2} = \frac{1}{D}\times$$
$$\times\Big\{\operatorname{Im}\varphi(z')\Big[\left(\operatorname{Re}Z^2(z') - \operatorname{Im}Z^2(z')\right)z'^2 - \left(\xi' + \operatorname{Re}\chi(z')\right)^2 + \left(\operatorname{Im}\chi(z')\right)^2 - \operatorname{Re}\varphi^2(z') + \operatorname{Im}\varphi^2(z')\Big] -$$
$$-2\operatorname{Re}\varphi(z')\Big[\operatorname{Re}Z(z')\operatorname{Im}Z(z')z'^2 - \operatorname{Im}\chi(z')\left(\xi' + \operatorname{Re}\chi(z')\right) - \operatorname{Re}\varphi(z')\operatorname{Im}\varphi(z')\Big]\Big\} \qquad (5)$$

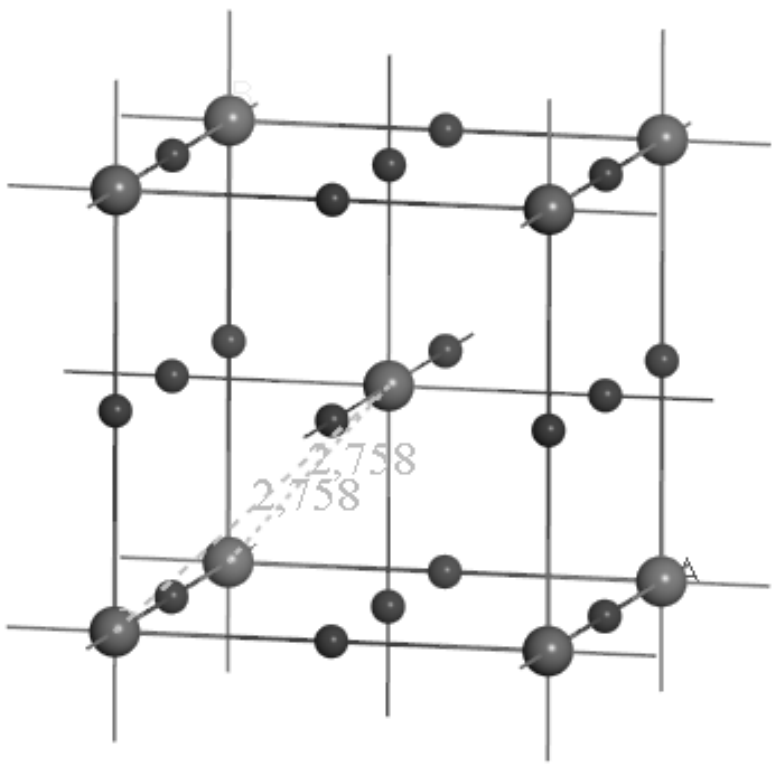

Fig.1. The SH3 structure investigated in the present paper is shown. Sulfur atoms are represented by a larger size balls. The Fig.1 is taken from [26].

Here

$$D = \Big[\left(\operatorname{Re}Z^2(z') - \operatorname{Im}Z^2(z')\right)z'^2 - \left(\xi' + \operatorname{Re}\chi(z')\right)^2 + \left(\operatorname{Im}\chi(z')\right)^2 - \operatorname{Re}\varphi^2(z') + \operatorname{Im}\varphi^2(z')\Big]^2 +$$
$$+4\Big[\operatorname{Re}Z(z')\operatorname{Im}Z(z')z'^2 - \operatorname{Im}\chi(z')\left(\xi' + \operatorname{Re}\chi(z')\right) - \operatorname{Re}\varphi(z')\operatorname{Im}\varphi(z')\Big]^2. \qquad (6)$$

Neglecting both the $\operatorname{Re}\Sigma(\xi,\omega)$ and $\operatorname{Im}\Sigma(\xi,\omega)$ dependence on $\xi$ we obtain omitting the small [27] quantities $\operatorname{Im}Z(\omega)$, $\operatorname{Re}\chi(\omega)$, $\operatorname{Im}\chi(\omega)$ in (5), (6), the following non-linear equation for the real part of the complex abnormal order parameter $\varphi$:

$$\operatorname{Re}\varphi(\omega) = -\frac{1}{\pi}P\int_0^{+\infty} dz'\left[K^{ph}(z',\omega) - K^{ph}(-z',\omega)\right]\times$$
$$\times\int_{-\mu}^{+\infty} d\xi' \frac{N_0(\xi')}{N_0(0)} \frac{2\operatorname{Re}\varphi^2(z')\operatorname{Im}\varphi(z')}{\left\{\left[\operatorname{Re}Z(z')\right]^2 z'^2 - (\xi')^2 - \operatorname{Re}\varphi^2(z') + \operatorname{Im}\varphi^2(z')\right\}^2 + 4\left[\operatorname{Re}\varphi(z')\operatorname{Im}\varphi(z')\right]^2} -$$
$$-\frac{1}{\pi}P\int_0^{+\infty} dz'\left[K^{ph}(z',\omega) - K^{ph}(-z',\omega)\right]\times \qquad (7)$$
$$\times\int_{-\mu}^{+\infty} d\xi' \frac{N_0(\xi')}{N_0(0)} \frac{\operatorname{Im}\varphi(z')\left\{\left[\operatorname{Re}Z(\xi',z')\right]^2 z'^2 - (\xi')^2 - \operatorname{Re}\varphi^2(z') + \operatorname{Im}\varphi^2(z')\right\}}{\left\{\left[\operatorname{Re}Z(z')\right]^2 z'^2 - (\xi')^2 - \operatorname{Re}\varphi^2(z') + \operatorname{Im}\varphi^2(z')\right\}^2 + 4\left[\operatorname{Re}\varphi(z')\operatorname{Im}\varphi(z')\right]^2}.$$

For the imaginary part of the complex anomalous order parameter $\varphi$ we similarly obtain the following expression:

$$\operatorname{Im}\varphi(\omega)=-\frac{1}{2}\int_{0}^{+\infty}dz'\{\ \alpha^2(|\omega-z'|)F(|\omega-z'|\omega-z')\left[cth\frac{(\omega-z')}{2T}+th\frac{z'}{2T}\right]sign(\omega-z')-$$
$$-\alpha^2(|\omega+z'|)F(|\omega+z'|)\left[cth\frac{(\omega+z')}{2T}-th\frac{z'}{2T}\right]sign(\omega+z')\ \}\times$$
$$\times\{\ \int_{-\mu}^{+\infty}d\xi'\frac{N_0(\xi')}{N_0(0)}\frac{2\operatorname{Re}\varphi^2(z')\operatorname{Im}\varphi(z')}{\left\{\left[\operatorname{Re}Z(z')\right]^2 z'^2-(\xi')^2-\operatorname{Re}\varphi^2(z')+\operatorname{Im}\varphi^2(z')\right\}^2+4\left[\operatorname{Re}\varphi(z')\operatorname{Im}\varphi(z')\right]^2}+ \tag{8}$$
$$+\int_{-\mu}^{+\infty}d\xi'\frac{N_0(\xi')}{N_0(0)}\frac{\operatorname{Im}\varphi(z')\left\{\left[\operatorname{Re}Z(\xi',z')\right]^2 z'^2-(\xi')^2-\operatorname{Re}\varphi^2(z')+\operatorname{Im}\varphi^2(z')\right\}}{\left\{\left[\operatorname{Re}Z(z')\right]^2 z'^2-(\xi')^2-\operatorname{Re}\varphi^2(z')+\operatorname{Im}\varphi^2(z')\right\}^2+4\left[\operatorname{Re}\varphi(z')\operatorname{Im}\varphi(z')\right]^2}\ \}.$$

Near Tc the product $\operatorname{Re}\varphi(z')\operatorname{Im}\varphi(z')$ tends to zero for all values of the argument, so that the expression

$$\frac{2\operatorname{Re}\varphi(z')\operatorname{Im}\varphi(z')}{\left\{\left[\operatorname{Re}Z(\xi',z')\right]^2 z'^2-(\xi')^2-\operatorname{Re}\varphi^2(z')+\operatorname{Im}\varphi^2(z')\right\}^2+4\left[\operatorname{Re}\varphi(z')\operatorname{Im}\varphi(z')\right]^2}$$ becomes delta

function $\pi\delta\left\{\left[\operatorname{Re}Z(z')^2 z'^2-(\xi')^2-\operatorname{Re}\varphi^2(z')+\operatorname{Im}\varphi^2(z')\right]^2\right\}\operatorname{sgn}\left[\operatorname{Re}\varphi(z')\operatorname{Im}\varphi(z')\right]$, whereas

the expression $\dfrac{\left\{\left[\operatorname{Re}Z(\xi',z')\right]^2 z'^2-(\xi')^2-\operatorname{Re}\varphi^2(z')+\operatorname{Im}\varphi^2(z')\right\}}{\left\{\left[\operatorname{Re}Z(\xi',z')\right]^2 z'^2-(\xi')^2-\operatorname{Re}\varphi^2(z')+\operatorname{Im}\varphi^2(z')\right\}^2+4\left[\operatorname{Re}\varphi(z')\operatorname{Im}\varphi(z')\right]^2}$

in the second term of (7), (8) simplifies to the following form:

$\left\{\operatorname{Re}Z(z')^2 z'^2-\operatorname{Re}\varphi^2(z')+\operatorname{Im}\varphi^2(z')-\xi'^2\right\}^{-1}$. As a result the nonlinear equations for the real part $\operatorname{Re}\varphi(\omega)$ of the order parameter after $\xi'$ integration with the account of the delta function properties is taken as follows:

$$\operatorname{Re}\varphi(\omega)=-P\int_{0}^{+\infty}dz'\left[K^{ph}(z',\omega)-K^{ph}(-z',\omega)\right]\frac{\operatorname{Re}\varphi(z')}{\sqrt{\operatorname{Re}^2 Z(z')z'^2-\operatorname{Re}\varphi^2(z')+\operatorname{Im}\varphi^2(z')}}\times$$
$$\times\frac{N_0\left(-\left|\operatorname{Re}^2 Z(z')z'^2-\operatorname{Re}\varphi^2(z')+\operatorname{Im}\varphi^2(z')\right|^{\frac{1}{2}}\right)+N_0\left(\left|\operatorname{Re}^2 Z(z')z'^2-\operatorname{Re}\varphi^2(z')+\operatorname{Im}\varphi^2(z')\right|^{\frac{1}{2}}\right)}{2N_0(0)}- \tag{9}$$
$$-\frac{1}{\pi}P\int_{0}^{+\infty}dz'\left[K^{ph}(z',\omega)-K^{ph}(-z',\omega)\right]\int_{-\mu}^{+\infty}d\xi'\frac{N_0(\xi')}{N_0(0)}\frac{\operatorname{Im}\varphi(z')}{\operatorname{Re}Z(z')^2 z'^2-\operatorname{Re}\varphi^2(z')+\operatorname{Im}\varphi^2(z')-\xi'^2}.$$

For the imaginary part $\operatorname{Im}\varphi(\omega)$ of the order parameter we obtain as a result of the $\xi$ integration taking into account the properties of the delta function the following expression:

$$\begin{gathered}
\operatorname{Im}\varphi(\omega)=-\frac{1}{2}\int_{0}^{+\infty}dz'\{\ \alpha^2(|\omega-z'|)F(|\omega-z'|)\left[cth\frac{(\omega-z')}{2\mathrm{T}}+th\frac{z'}{2\mathrm{T}}\right]sign(\omega-z')-\\
-\alpha^2(|\omega+z'|)F(|\omega+z'|)\left[cth\frac{(\omega+z')}{2\mathrm{T}}-th\frac{z'}{2\mathrm{T}}\right]sign(\omega+z')\ \}\left\{-\frac{\pi\operatorname{Re}\varphi(z')}{\sqrt{\operatorname{Re}^2 Z(z')z'^2-\operatorname{Re}\varphi^2(z')+\operatorname{Im}\varphi^2(z')}}\times\right.\\
\times\left[\frac{N_0\left(-\left|\operatorname{Re}^2 Z(z')z'^2-\operatorname{Re}\varphi^2(z')+\operatorname{Im}\varphi^2(z')\right|^{\frac{1}{2}}\right)}{2N_0(0)}+\frac{N_0\left(\left|\operatorname{Re}^2 Z(z')z'^2-\operatorname{Re}\varphi^2(z')+\operatorname{Im}\varphi^2(z')\right|^{\frac{1}{2}}\right)}{2N_0(0)}\right]+\\
\left.-P\int_{-\mu}^{+\infty}d\xi'\frac{N_0(\xi')}{N_0(0)}\frac{\operatorname{Im}\varphi(z')}{\left[\operatorname{Re}Z(\xi',z')\right]^2 z'^2-(\xi')^2-\operatorname{Re}\varphi^2(z')+\operatorname{Im}\varphi^2(z')}\ \right\}.
\end{gathered}\tag{10}$$

The right hand side of equations (9) - (10) for the complex order parameter contains previously not taken into account $[12-24]$ contributions proportional to $\operatorname{Im}\varphi(z')$. The order parameter will be written in the following form $\varphi(\omega)=\Delta(\omega)|Z(\omega)|$, $|Z(z')|=\left(\operatorname{Re}^2 Z(z')+\operatorname{Im}^2 Z(z')\right)^{\frac{1}{2}}$ , the $z'$ integral in (9), (10) is taken as a principal value, that is marked with the P mark, the negative $-\left|\operatorname{Re}^2 Z(z')z'^2-\operatorname{Re}\varphi^2(z')+\operatorname{Im}\varphi^2(z')\right|^{\frac{1}{2}}$ value can not be less than $-\mu$, so that the integration over $z'$ at negative $z'$ breaks provided $\left|\operatorname{Re}^2 Z(z')z'^2-\operatorname{Re}\varphi^2(z')+\operatorname{Im}\varphi^2(z')\right|^{\frac{1}{2}}=\mu$.The integrand is equal to zero under such $z'$ that $\operatorname{Re}^2 Z(z')z'^2-\operatorname{Re}\varphi^2(z')+\operatorname{Im}\varphi^2(z')<0$.The root will assume to be positive $\sqrt{\operatorname{Re}^2 Z(z')z'^2-\operatorname{Re}\varphi^2(z')+\operatorname{Im}\varphi^2(z')}\geq 0$ for any $z'$ sign. Assuming the constancy of the "bare" electron density of states $N_0(\omega)$, we can pass from a system of equations (9) - (10) to the conventional system of Éliashberg equations $[10-22,24]$ in which the final width of the electron band, the pairing outside the Fermi-surface, the variability of the electronic density of states and the effects of electron-hole non-equivalence are neglected. In the wide-band materials such as the hydrogen sulphide, the logarithmic term in the last two equations with the high accuracy can be set equal to zero, so that the system of equations for the order parameter takes the form:

$$\operatorname{Re}\varphi(\omega) = -P\int_0^{+\infty} dz'\left[K^{ph}(z',\omega) - K^{ph}(-z',\omega)\right]\frac{\operatorname{Re}\varphi(z')}{\sqrt{\operatorname{Re}^2 Z(z')z'^2 - \operatorname{Re}\varphi^2(z') + \operatorname{Im}\varphi^2(z')}}\times$$
$$\times\frac{N_0\left(-\left|\operatorname{Re}^2 Z(z')z'^2 - \operatorname{Re}\varphi^2(z') + \operatorname{Im}\varphi^2(z')\right|^{\frac{1}{2}}\right) + N_0\left(\left|\operatorname{Re}^2 Z(z')z'^2 - \operatorname{Re}\varphi^2(z') + \operatorname{Im}\varphi^2(z')\right|^{\frac{1}{2}}\right)}{2N_0(0)}.$$

(11)

$$\operatorname{Im}\varphi(\omega) = -\frac{1}{2}\int_0^{+\infty} dz'\{\ \alpha^2(|\omega - z'|)F(|\omega - z'|)\left[cth\frac{(\omega - z')}{2\mathrm{T}} + th\frac{z'}{2\mathrm{T}}\right]sign(\omega - z') -$$
$$-\alpha^2(|\omega + z'|)F(|\omega + z'|)\left[cth\frac{(\omega + z')}{2\mathrm{T}} - th\frac{z'}{2\mathrm{T}}\right]sign(\omega + z')\ \}\frac{\pi\operatorname{Re}\varphi(z')}{\sqrt{\operatorname{Re}^2 Z(z')z'^2 - \operatorname{Re}\varphi^2(z') + \operatorname{Im}\varphi^2(z')}}\times$$
$$\times\left[\frac{N_0\left(-\left|\operatorname{Re}^2 Z(z')z'^2 - \operatorname{Re}\varphi^2(z') + \operatorname{Im}\varphi^2(z')\right|^{\frac{1}{2}}\right)}{2N_0(0)} + \frac{N_0\left(\left|\operatorname{Re}^2 Z(z')z'^2 - \operatorname{Re}\varphi^2(z') + \operatorname{Im}\varphi^2(z')\right|^{\frac{1}{2}}\right)}{2N_0(0)}\right].$$

(12)

When assuming constant electronic density of states $N_0\left(\pm\left|\operatorname{Re}^2 Z(z')z'^2 - \operatorname{Re}\varphi^2(z') + \operatorname{Im}\varphi^2(z')\right|^{\frac{1}{2}}\right) = const$ the equations for the complex order parameter near Tc are simplified to the following form:

$$\operatorname{Re}\varphi(\omega) = -P\int_0^{+\infty} dz'\left[K^{ph}(z',\omega) - K^{ph}(-z',\omega)\right]\frac{\operatorname{Re}\varphi(z')}{\sqrt{\operatorname{Re}^2 Z(z')z'^2 - \operatorname{Re}\varphi^2(z') + \operatorname{Im}\varphi^2(z')}},$$

( 13)

$$\operatorname{Im}\varphi(\omega) = -\frac{1}{2}\int_0^{+\infty} dz'\{\ \alpha^2(|\omega - z'|)F(|\omega - z'|)\left[cth\frac{(\omega - z')}{2\mathrm{T}} + th\frac{z'}{2\mathrm{T}}\right]sign(\omega - z') -$$
$$-\alpha^2(|\omega + z'|)F(|\omega + z'|)\left[cth\frac{(\omega + z')}{2\mathrm{T}} - th\frac{z'}{2\mathrm{T}}\right]sign(\omega + z')\ \}\frac{\pi\operatorname{Re}\varphi(z')}{2\sqrt{\operatorname{Re}^2 Z(z')z'^2 - \operatorname{Re}\varphi^2(z') + \operatorname{Im}\varphi^2(z')}}.$$

(14)

Usually already extremely oversimplified equations (13) - (14) for the superconducting order parameter are solved neglecting the imaginary part of the order parameter, that leads to the next standard [17] form as follows:

$$\operatorname{Re}\varphi(\omega) = -P\int_{0}^{+\infty} dz'\left[K^{ph}(z',\omega) - K^{ph}(-z',\omega)\right]\frac{\operatorname{Re}\varphi(z')}{\sqrt{\operatorname{Re}^2 Z(z')z'^2 - \operatorname{Re}\varphi^2(z')}}. \tag{15}$$

From the above it is clear that the use of such a simplified form of the Éliashberg equations can not qualify for a quantitative description of the superconducting transition temperature Tc in the specific materials.

## 2. High $T_c$ in the hydrogen sulfide as a result of the variability of the density of electron states in the band

In this paper we determine Tc and the frequency behavior of the complex order parameter at different temperatures by solving the Éliashberg equations in the form of a non-linear system of equations (11) - (12) for the complex order parameter in hydrogen sulfide $SH_3$ phase at the pressure of 225 GPa with the full account of the alternative character of the electronic density of states.

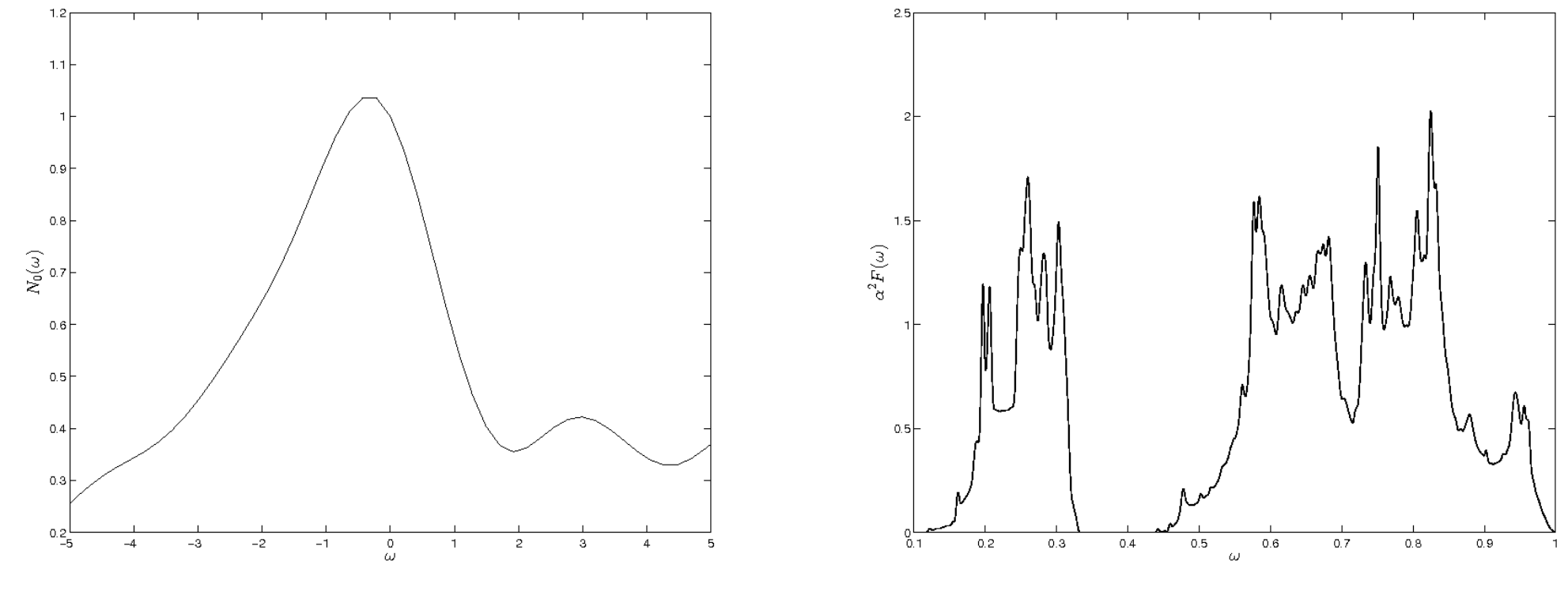

Fig. 2. a. Dimensionless "bare" total density of $SH_3$ hydrogen sulfide electronic states at the pressure 225 GPa [26].; b. The spectral function of the electron-phonon interaction $\alpha^2(\omega)F(\omega)$ [26] in $SH_3$ hydrogen sulfide at the pressure 225 GPa. The frequency $\omega$ is expressed in dimensionless units (as a fraction of the maximum frequency of the phonon spectrum).

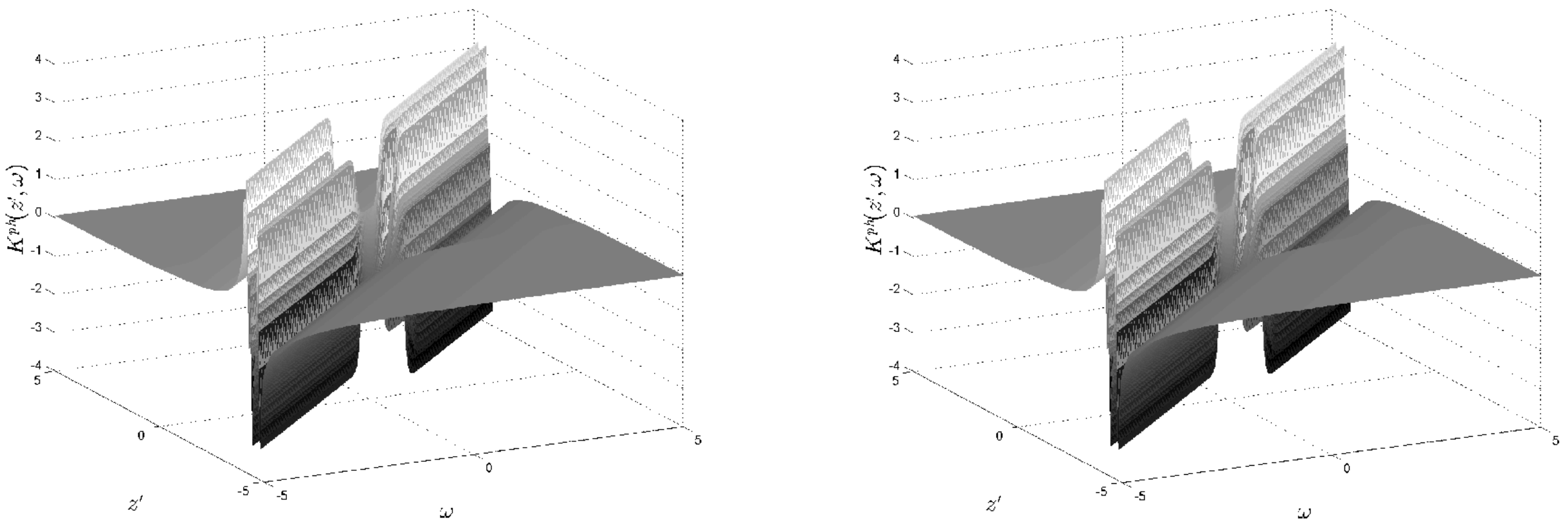

Fig.3 K-function (4) of the $SH_3$ metal hydrogen sulphide phase as a function of two dimensionless parameters $z'$ and $\omega$ at different temperatures a. T=175K, b. T=205K.

The functional $\mathrm{Re}\,Z(\omega,T)$, $N_0(\omega,T)$ dependencies at different temperatures, contained in the expressions (11) - (12), were calculated using the formalism developed in $[27]$. The frequency dependence of both the electron mass operator $\mathrm{Re}\,Z(\omega,T), \mathrm{Im}\,Z(\omega,T)$ renormalization and of the terms $\mathrm{Re}\,\chi(\omega)$, $\mathrm{Im}\,\chi(\omega)$ usually called the " complex renormalization of the chemical potential ," are all presented in Fig.4:

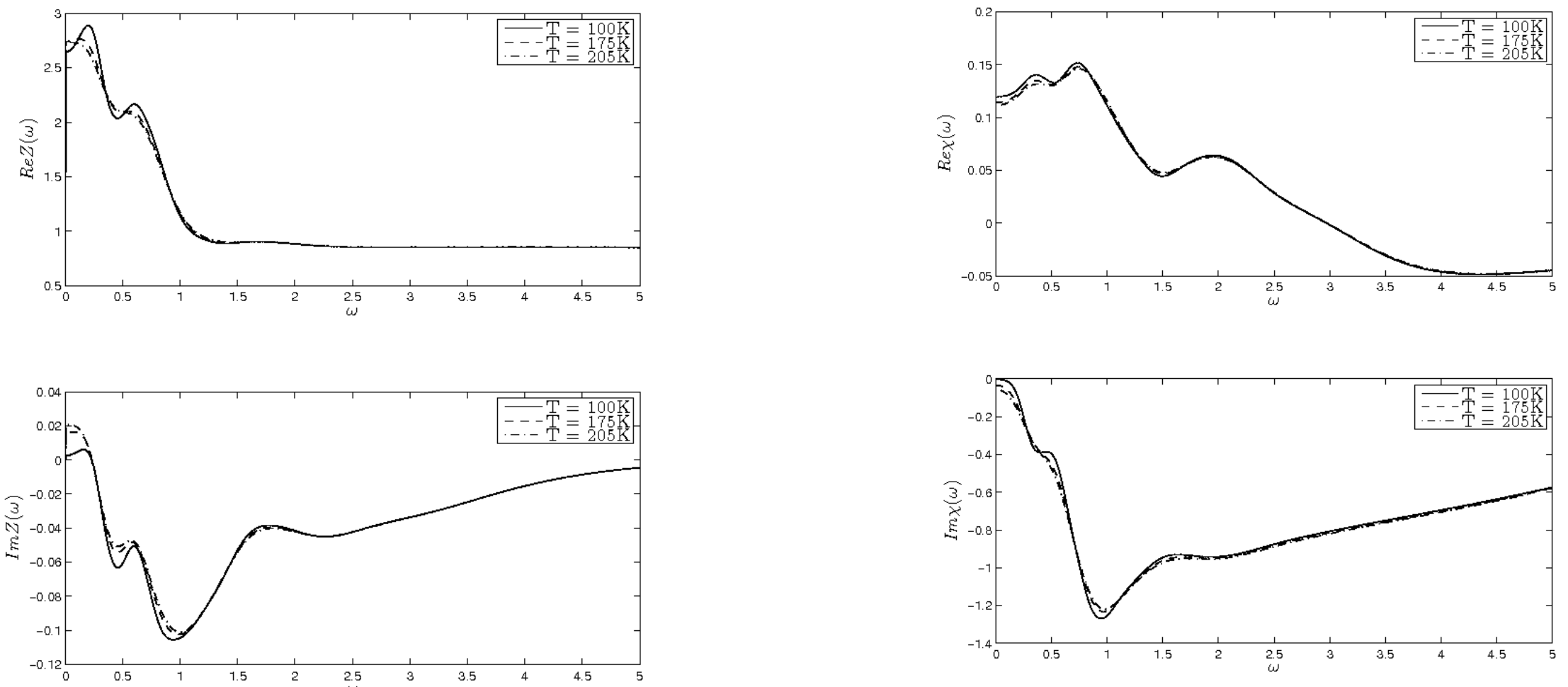

Fig. 4. The reconstructed conductivity band parameters of the $SH_3$ metal hydrogen sulphide phase. a. The real part $\mathrm{Re}\, Z(\omega)$ of the electron mass renormalization $Z(\omega)$, b. The imaginary part $\mathrm{Im}\, Z(\omega)$ of the renormalization of the electron mass in the self-energy part of the electron Green function. Renormalized with the electron-phonon interaction the real part $\mathrm{Re}\,\chi(\omega)$ of the renormalization of the chemical potential. Renormalized with the electron-phonon interaction the imaginary part of the renormalization of the chemical potential $\mathrm{Im}\,\chi(\omega)$. Frequency $\omega$ is expressed in the dimensionless units (as a fraction of the maximum frequency 0.214 eV of the phonon spectrum for this hydrogen sulfide phase). All results were obtained for the pressure P = 225 GPa and the temperature T = 200K.

The system of equations (11), (12) is solved by using an iterative method based on the frequency behavior of the spectral function of the EP interaction (function Éliashberg) $\alpha^2 F(z)$ [27] for $SH_3$ hydrogen sulfide phase at at the pressure P = 225 GPa. It was found that the process of the convergence of the solution for the real part of the order parameter $\mathrm{Re}\,\varphi(\omega)$ in solving the system of equations (11), (12) is set when the number of several tens of iterations is reached. At T = 180K and atT = 300K $\mathrm{Re}\,\varphi(\omega)$ as well as $\mathrm{Im}\,\varphi(\omega)$ tend to zero with increasing number of iterations thus indicating no effect of superconductivity at these temperature. In this case, however, the order parameter preserves the characteristic structure of the superconducting state decreasing with increasing number of iterations. Equations (11), (12) below the $T_c$ temperature have a set of three solutions namely: $\mathrm{Re}\,\varphi(\omega)$ and $\mathrm{Im}\,\varphi(\omega)$, $-\mathrm{Re}\,\varphi(\omega)$ and $-\mathrm{Im}\,\varphi(\omega)$ and in the case of the presence of superconductivity additionally the zero unstable solution. In the numerical solution of equations (11) - (12) on the real axis the solution before the establishment of the zero solution undergoes multiple rebuilds from "negative" to the "positive" solutions. An additional difficulty in solving the equations (11) - (12) is a numerical integration of improper integrals with divergences appearing in these equations. The behavior of the real part of the order parameter $\mathrm{Re}\,\varphi(\omega)$ as well as the behavior of the imaginary part of the order parameter $\mathrm{Im}\,\varphi(\omega)$ at the temperatures T = 175K, 180K, 300K is shown in Fig. 5-7. The order parameter solution for the resulting value of $T_c$ = 177K is not presented here due to the vanishing order parameter values at this temperature.

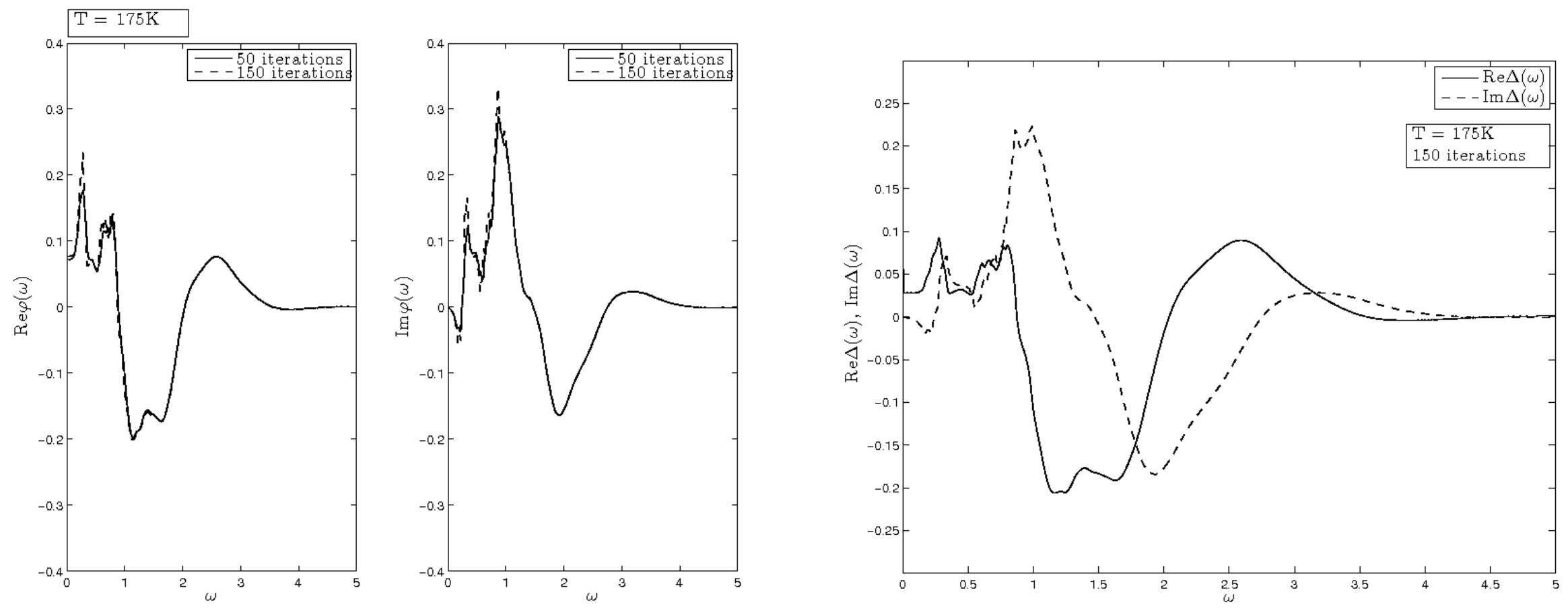

Fig. 5. The frequency dependence of the steady solution for a. the real part $\mathrm{Re}\,\varphi(\omega)$ and the imaginary part $\mathrm{Im}\,\varphi(\omega)$; b. the real part $\mathrm{Re}\,\Delta(\omega)$ and the imaginary part $\mathrm{Im}\,\Delta(\omega)$ of the $SH_3$ hydrogen sulfide phase order parameter at T = 175K and at the pressure P = 225 GPa. The frequency is expressed in the dimensionless units, corresponding to the limiting frequency $0.234 meV$ of the phonon spectrum.

The imaginary part $\mathrm{Im}\,\Delta(\omega)$ of the order parameter at low frequency is negative, while at the value of the dimensionless frequency equal to 0.23, Fig.5.b. , the imaginary part $\mathrm{Im}\,\Delta(\omega)$ value becomes positive. Thus, we set the value of the energy gap in the $SH_3$ sulfide phase, which turned out to be $0.23 \times 0.234 meV$, that is, 592 degrees. Fig. 6 shows the process of the $\varphi(\omega)$ vanishing with the increasing number of iterations of the solution for the complex $\varphi(\omega)$ function at the temperature T = 180K, thus indicating that $T_c < 180K$.

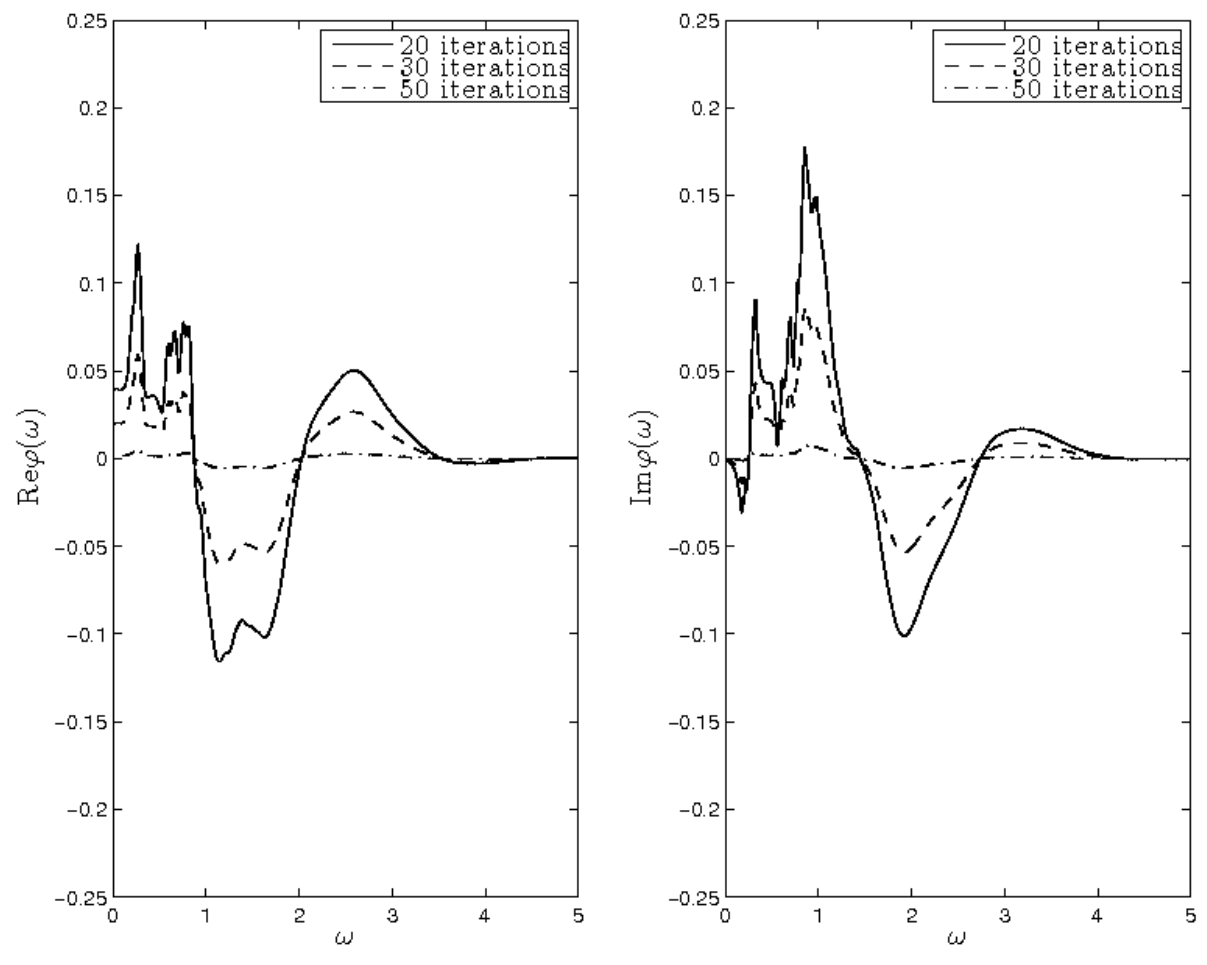

Fig. 6. The dependence of the solution on the iteration number for the real part $\mathrm{Re}\,\varphi(\omega)$ and for the imaginary part $\mathrm{Im}\,\varphi(\omega)$ of the hydrogen sulphide $SH_3$ phase order parameter at T = 180K at the pressure P = 225 GPa. The frequency is expressed in the dimensionless units, corresponding to the limiting frequency $0.234meV$ of the phonon spectrum .

Fig. 7 shows the frequency dependence at the number 100 of iterations of the very small quantities $\mathrm{Re}\varphi(\omega)$, $\mathrm{Im}\varphi(\omega)$ at the temperature T = 300K. From the Fig. 7 it is clearly seen that the order parameter $\mathrm{Re}\varphi(\omega)$, $\mathrm{Im}\varphi(\omega)$ magnitude in the decrease with increasing number of iterations even at T = 300K retains characteristic functional dependency of the hydrogen sulfide order parameter in the superconducting state. At the same time the rough approximations of 1 for $\mathrm{Re}\varphi(\omega)$ and of 0 for $\mathrm{Im}\varphi(\omega)$ are used as the initial conditions introducing in the solution no functional dependency on the frequency $\omega$.

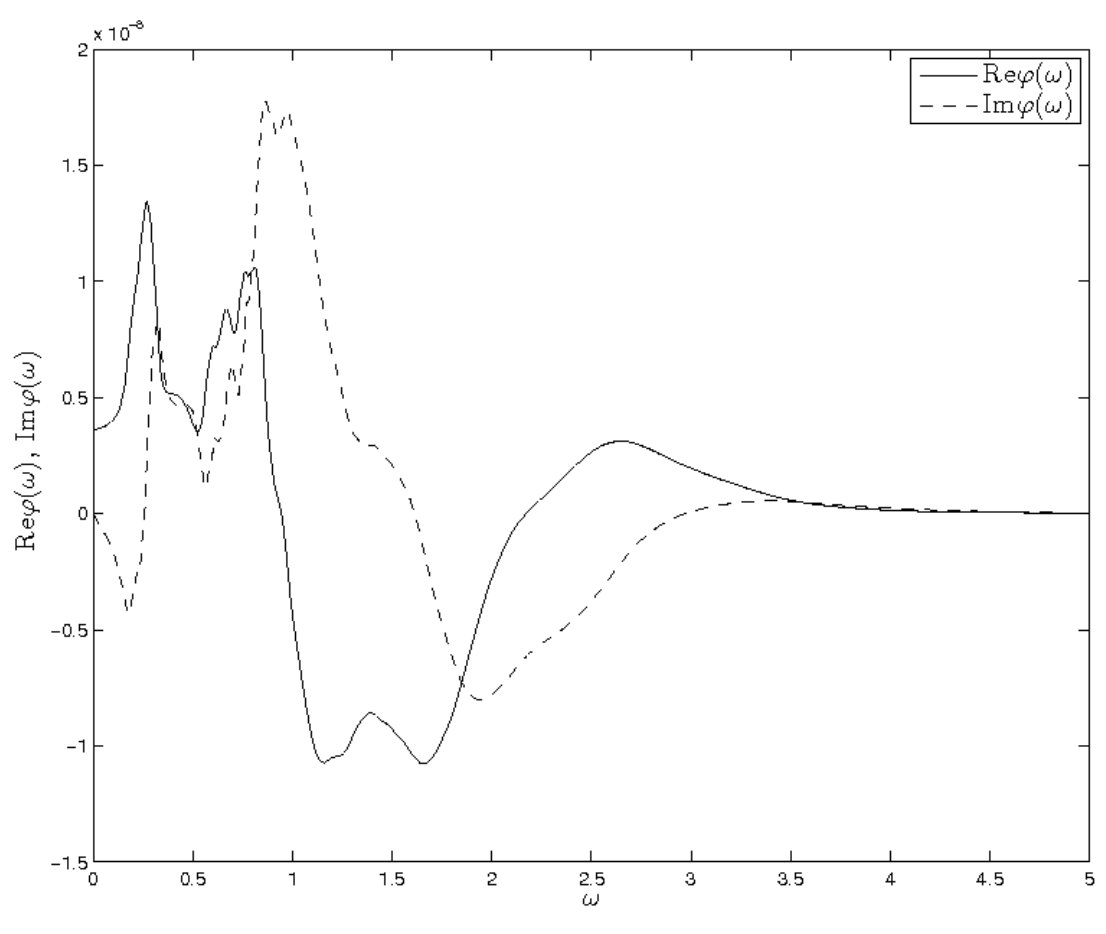

Fig. 7. The frequency dependence of the order parameter solutions at the hundred iterations for the real part $\mathrm{Re}\,\varphi(\omega)$ and the imaginary part $\mathrm{Im}\,\varphi(\omega)$ of the hydrogen sulphide $SH_3$ phase at T = 300 K at the pressure P = 225 GPa. The frequency is expressed in dimensionless units, corresponding to the limiting frequency $0.234meV$ of the phonon spectrum.

Fig. 8 shows the dependence of the order parameter's steady-state solutions on the temperature at the temperatures below the critical temperature Tc.

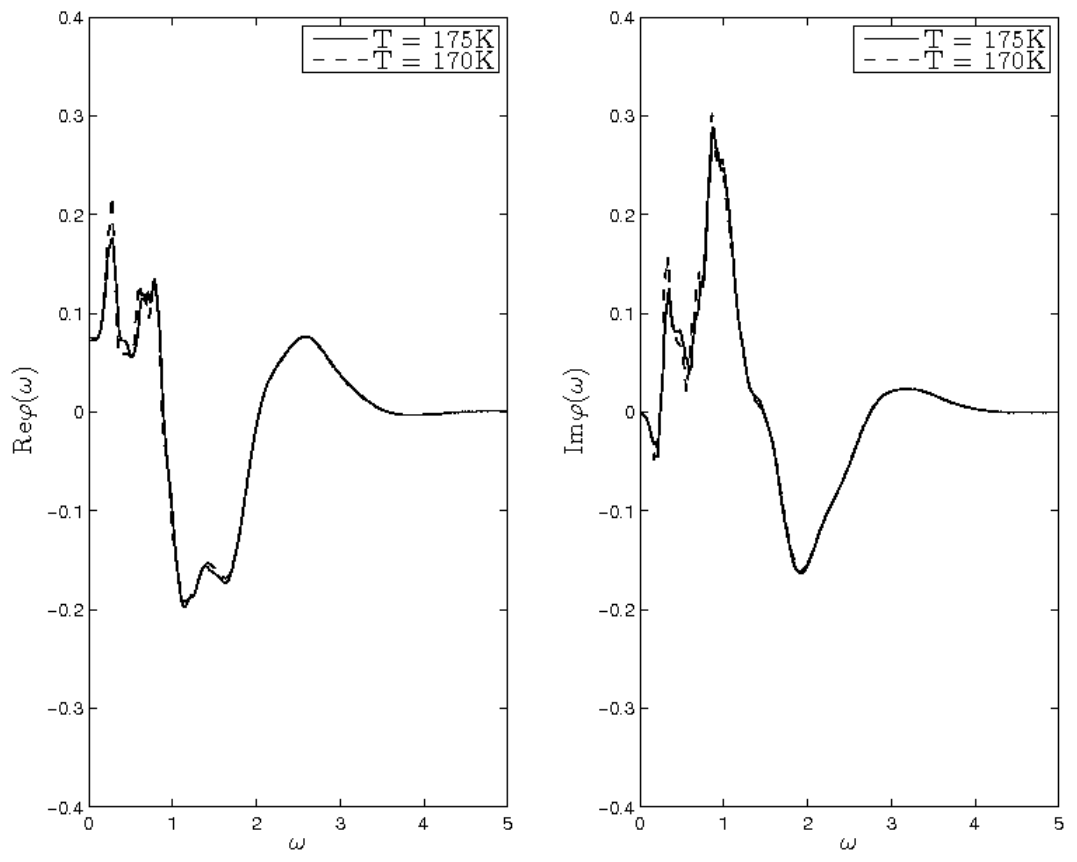

Fig. 8. Temperature dependence of the steady-state solutions (50 iterations) for the order parameter real part $\mathrm{Re}\,\varphi(\omega)$ and the imaginary part $\mathrm{Im}\,\varphi(\omega)$ for the hydrogen sulfide $SH_3$ phase at the pressure P = 225 GPa. The frequency is expressed in the dimensionless units, corresponding to the limiting frequency $0.234 meV$ of the phonon spectrum.

## 3. Conclusions

Analyzing the results and summing everything written before, we arrive at the following conclusions: 1. Éliashberg theory is generalized to account for the variable nature of the electron density of states. 2. The mathematical method for the solving of Éliashberg equations on the real axis is developed. 3. The generalized Éliashberg equations with the variable nature of the electron density of states are solved. The quantitative agreement with the experiment for the hydrogen sulphide $SH_3$ phase is achieved. The fact of the very slow convergence of the solutions of Éliashberg equations with increasing iteration number is established. 4. The frequency dependence as well as the fine structure of both the real part $\mathrm{Re}\,\varphi(\omega)$ and the imaginary part $\mathrm{Im}\,\varphi(\omega)$ of the order parameter corresponding to the selected $SH_3$ sulfide phase at the temperatures T = 175K , T = 180K, T = 300K is obtained. The variation of the frequency dependence of the order parameter with the temperature is presented. 5. The energy gap magnitude of the $SH_3$ sulfide phase is found to have the value 592 degrees. 6. It is shown that at the temperatures above the critical temperature the order parameter very slowly tends to zero with increasing the number of iterations, maintaining the characteristic functional behavior of the superconducting state on the frequency. 7. The method of solving of the Éliashberg equations on a set of discrete points on the imaginary axis, faced with the problem of the convergence of the solution at low order of the discrete matrix, does not accurately reproduce the dependence of the order parameter on the frequency after the analytic continuation of the order parameter to the real frequency axis compared to the method of solving of Éliashberg equations on the real axis. It is shown that at the temperatures above the critical one the order parameter is very slowly tending to zero with increasing iteration number, maintaining the characteristic functional behavior of the superconducting state on the frequency. 8. All the calculations were carried out ab-initio. The work is not using any assumptions or any fitting parameters. The entire treatment was carried out on the real axis so as to be able to explore the frequency behavior of the order parameter simultaneously with the $T_c$ calculation without the analytic continuation procedure. We got $T_c = 177K$ coinsiding with the experimental [2] $T_c$ value in the hydrogen sulfide at the pressure 225GPa. At the temperature $T = 180K > T_c$ and even at the room temperature T = 300 K, when the equation for the order parameter leads to an extremely small maximum values of the order parameter $\mathrm{Re}\,\varphi,\ \mathrm{Im}\,\varphi \sim 10^{-8}$ on the hundredth iteration, the order parameter frequency dependence is similar with the order parameter dependence on frequency for the superconducting state. 9. Three factors influence the $T_c$ of the EP system in the critical mode, namely: the variable nature of the electron density of states $N_0(\varepsilon)$, the specific properties of a substance depending on the electron mass renormalization $\mathrm{Re}\,Z(z)$, the damping $\mathrm{Im}\,Z(z)$ of the electrons, both the real $\mathrm{Re}\,\chi(\omega)$ and the imaginary $\mathrm{Im}\,\chi(\omega)$ components of the renormalization of the chemical potential, as well as the novel contributions $\sim \mathrm{Im}\,\Delta(\omega)$ to the Éliashberg equations which were not taken into account in the formalism of the previous works $[12-24]$.The neglect of the terms proportional to $\operatorname{Im}\Delta(\omega)$ leads to the violation of the Kramers-Kronig relations for the imaginary and the real part of the order parameter in the Éliashberg equations. 10. It is of critical importance to take into account the variability of the density of electron states in the conduction band for the high $T_c$ appearance in the EP system. The accounting of the electron density of states variability in the band leads to the possibility of the pairing of electrons in the entire Fermi – volume in contrast to the usually considered pairing within the layer with $\omega_D$ thickness at the Fermi surface.11. The Coulomb pseudopotential of electrons in the hydrogen sulfide leads to the inessential $T_c$ reduction. From the present consideration, given the considerable 592 degrees value of the superconducting gap and taking into account the conservation of the diminishing order parameter structure for the temperature larger $T_c$, it becomes clear that the hydrogen sulfide EP system even at the temperatures above the critical point remains in the "quasi superconducting " state. From this it follows that the $T_c$ value in the EP system can be greatly improved as compared to the experimentally determined [1,2] $T_c$ value in the hydrogen sulfide by changing the pressure and with the selection of optimal $\operatorname{Re}Z(\omega), \operatorname{Im}Z(\omega)$, $\operatorname{Re}\chi(\omega)$, $\operatorname{Im}\chi(\omega)$ behavior, along with the optimal behavior of the electron density of states $N_0(\varepsilon)$ with a weak electron-hole non-equivalence and a moderate value of the EP interaction. It is also possible that the high temperature superconductivity can be achieved with the influence on the EP system being in a "quasi superconducting " state with the unknown specific weak perturbation the nature of which is still to be established.

The authors thank Yu. Kagan for the deep and stimulating discussion of this work. The study was supported by a grant from the Russian Science Foundation (project №14-11-00258).

[a] Mailing address: *EAMazur@mephi.ru*